\documentclass{article}
\usepackage{spconf,amsmath,graphicx,multirow,bm,float}

\usepackage{hyperref}

\title{CONFIDENCE ESTIMATION FOR BLACK BOX AUTOMATIC SPEECH RECOGNITION\\ SYSTEMS USING LATTICE RECURRENT NEURAL NETWORKS}
\name{A.~Kastanos$^{\star}$, A.~Ragni$^{\dagger,\ddagger}$\thanks{$\ddagger$ Supported in part by the
Office of the Director of National Intelligence (ODNI), Intelligence Advanced Research Projects Activity (IARPA), via Air Force Research Laboratory (AFRL) contract \# FA8650-17-C-9117. The views and conclusions contained herein are those of the authors and should not be interpreted as necessarily representing the official policies, either expressed or implied, of ODNI, IARPA, AFRL or the U.S. Government. The U.S. Government is authorised to reproduce and distribute reprints for governmental purposes notwithstanding any copyright annotation therein.}, M.~J.~F.~Gales$^{\star,\ddagger}$}
\address{$^{\star}$Department of Engineering, University of Cambridge, Trumpington Street, Cambridge CB2 1PZ, UK\\
$^{\dagger}$Department of Computer Science, University of Sheffield, 211 Portobello, Sheffield S1 4DP, UK\\
{\small\tt ak2132@cam.ac.uk, a.ragni@sheffield.ac.uk, mjfg@eng.cam.ac.uk}}
\begin{document}
\ninept
\maketitle
\begin{abstract}
Recently, there has been growth in providers of speech transcription services enabling others to leverage technology they would not normally be able to use. As a result, speech-enabled solutions have become commonplace. Their success critically relies on the quality, accuracy, and reliability of the underlying speech transcription systems. Those black box systems, however, offer limited means for quality control as only word sequences are typically available. This paper examines this limited resource scenario for confidence estimation, a measure commonly used to assess transcription reliability. In particular, it explores what other sources of word and sub-word level information available in the transcription process could be used to improve confidence scores. To encode all such information this paper extends lattice recurrent neural networks to handle sub-words. Experimental results using the IARPA OpenKWS 2016 evaluation system show that the use of additional information yields significant gains in confidence estimation accuracy. The implementation for this model can be found online \footnote{\scriptsize\url{https://github.com/alecokas/BiLatticeRNN-Confidence}}.
\end{abstract}
\section{Introduction}
\label{sec:intro}
Automatic Speech Recognition (ASR) has seen a surge in interest as speech enabled devices continue to proliferate the consumer market. From dedicated voice activated assistants, such as Amazon Alexa \cite{Ram2018aS} and Google Home \cite{Li2017aS}, to virtual assistants embedded in general purpose devices, such as Microsoft Cortana and Apple Siri, speech recognition is fast becoming a mainstream medium for interacting with technology. Though access to the underlying ASR technology has become easier \cite{htkbook2015aS, povey2011aS}, more and more ASR systems are purchased as {\em black box} models in the sense that the internal state of the system is inaccessible to the user. This is particularly common in cloud-based solutions where transcriptions are often served via an application programming interface (API). The usability of these black box ASR technologies is determined by their ability to produce a correct transcription for a given audio signal. Though efforts are made to ensure they can operate over a wide variety of conditions \cite{Li2017aS}, it is hard to guarantee high transcription quality for all possible scenarios. As a result, error mitigation strategies have become important. 

Confidence scores provide a mechanism to mitigate error-prone ASR systems by presenting a measure of uncertainty for intelligent post-processing modules \cite{hazen2002recognition, jiang2005confidence}. These scores also find applications within upstream tasks, such as speaker adaptation \cite{uebel2001speaker} and semi-supervised training \cite{chan2004improving}, and downstream tasks, such as machine translation \cite{Saleem2004aS} and information retrieval \cite{zbib2019aS}.  In the simplest case, confidence scores are posterior probabilities derived during the normal decoding process \cite{mangu1999finding, Evermann2000aL}. These scores are often the only uncertainty information provided even though rich graph representations, which encode multiple hypotheses at the sub-word level, are being generated during decoding. Posterior probabilities, however, are known to over-estimate confidence \cite{Evermann2000aL}. Though prior work exists on improving confidence estimation \cite{Evermann2000aL, Weintraub1997aS, Seigel2011aS, Kalgaonkar2015aS, DelAgua2018aS, ragni2018confidence, Li_2019}, no one has examined the impact of readily available information on the ability of black box ASR users to improve confidence estimates. 

This paper examines confidence estimation when limited information is available. In particular, it shows that significantly more accurate estimates can be obtained if additional information is propagated by these black box systems. In order to encode complex and rich graph representations, which combine information supplied by the black box system and the user, this paper extends bi-directional lattice recurrent neural networks (BiLatRNN) \cite{Li_2019} from the word level to include sub-word level features. Two attention-based approaches for handling variable length sub-word sequences are proposed. The more complex bi-directional encoder approach is found to be more accurate than the simpler self-attention approach \cite{Vaswani2017aS}.

The rest of this paper is organised as follows. Section~\ref{sec:blackbox} discusses standard representations of information within black box ASR systems, which includes graphs encoding alternative transcriptions and sub-word units. A neural network approach for encoding word level graphs is discussed in Section~\ref{sec:latrnn}. Section~\ref{sec:featues} describes standard word level features as well as introduces two approaches for encoding sub-word information. Experimental results are presented in Section~\ref{sec:exp}. The conclusions drawn from this work are given in Section~\ref{sec:conclusion}.

\section{Black box ASR systems}
\label{sec:blackbox}
A black box ASR system is a solution provided by an external company or individual for the task of speech transcription. Such solutions are particularly popular among early-stage companies or those not primarily focused on ASR. Based on their physical location, black box ASR systems can be divided into on-premise and cloud-based. On-premise solutions physically operate on user premises to support applications with certain restrictions on security and latency of transcription, while cloud-based solutions delegate transcription to a remote server that may be optimised to offer higher accuracy. 

\begin{figure}[htbp]
    \centering
    \begin{tabular}{cc}
    \includegraphics[width=3.5cm]{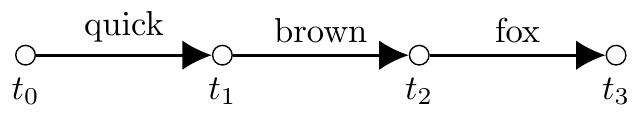} & \includegraphics[width=4.5cm]{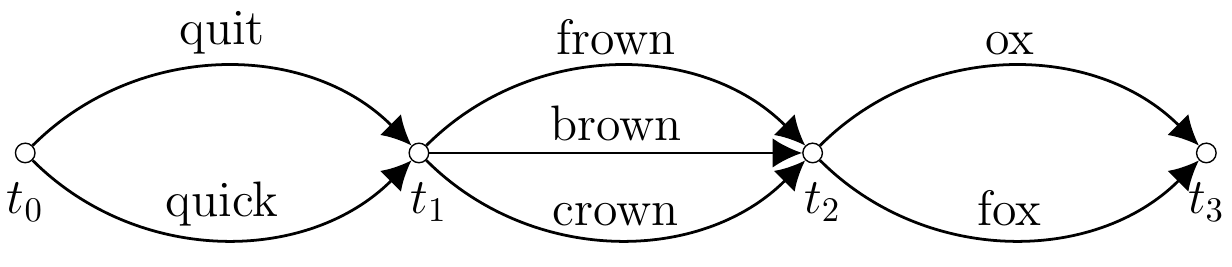}\\
    {\small (a) one-best sequence} & {\small (b) confusion network}\\
    \end{tabular}
    \begin{tabular}{c}
    \includegraphics[width=5cm]{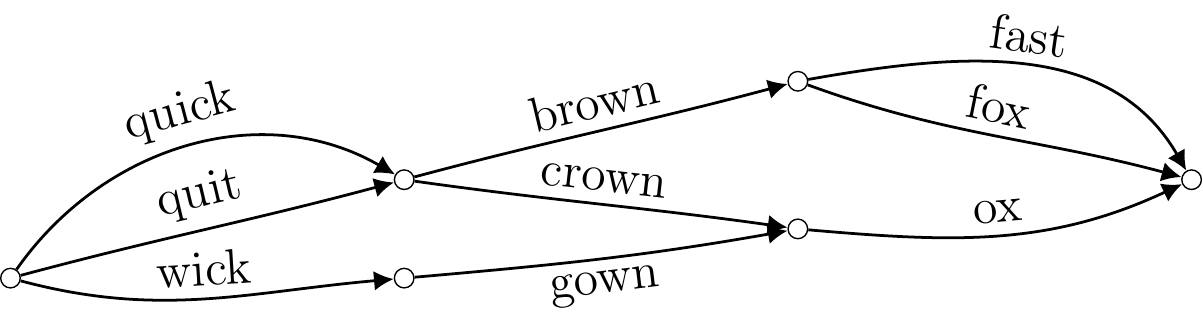}\\
    {\small (c) lattice}\\
    \end{tabular}
    \caption{Standard speech recognition output representations}
    \label{fig:one-best}
\end{figure}

Despite many advances in the speech recognition field to speed up decoding, and in the digital communication field to offer ultra-fast data transmission, black box ASR systems continue to provide only a very limited amount of information about the transcription process even when located on the user premises. Such information typically contains start and end times of the first and last transcribed word and the complete, one-best, word sequence. Figure \ref{fig:one-best}~(a) provides a graphical representation for the one-best word sequence corresponding to the hypothetical utterance \emph{quick brown fox}. Despite their restrictive nature, one-best sequences are the \emph{de facto} standard output provided by commercial systems and are commonly used by downstream applications in natural language processing.

Since word sequences alone carry no information about how certain the black box ASR system is, downstream applications have limited means for addressing transcription errors. The addition of simple features, such as word posterior probabilities and durations, provides the potential for a significantly better error mitigation mechanism to be devised. Other potentially useful characteristics include the word confusions found in confusion or consensus \cite{mangu1999finding, Evermann2000aL} networks (CN) illustrated by Figure \ref{fig:one-best}~(b). Such networks are typically generated by the black box ASR as a part of one-best generation process. CNs are a type of linear directed acyclical graph (DAG) which provides information pertaining to the most likely candidate transcriptions. Not only do the word confusions in CNs provide alternative word hypotheses, but they also provide an indication of the confidence in the prediction. Such rich and compact output representations have been found to be crucial for developing accurate downstream applications \cite{Le2014aS, zbib2019aS}, and are expected \cite{Li_2019} to benefit confidence estimation for one-best word sequences.

CNs are normally derived from a more general DAG representation produced during decoding. These graphs, or {\em lattices}, illustrated by Figure~\ref{fig:one-best}~(c), encode a wealth of information coming from acoustic, language and pronunciation models. The clustering process behind CN construction combines multiple, not necessary precisely overlapping in time, lattice arcs to yield one CN arc, thus losing the individual sources of information. Those sources can be linked to confusion network arcs and leveraged for confidence estimation if lattices were made available by black box ASR developers. 

\section{Lattice recurrent neural networks}
\label{sec:latrnn}
\begin{figure}[htbp]
    \centering
    \includegraphics[width=7cm]{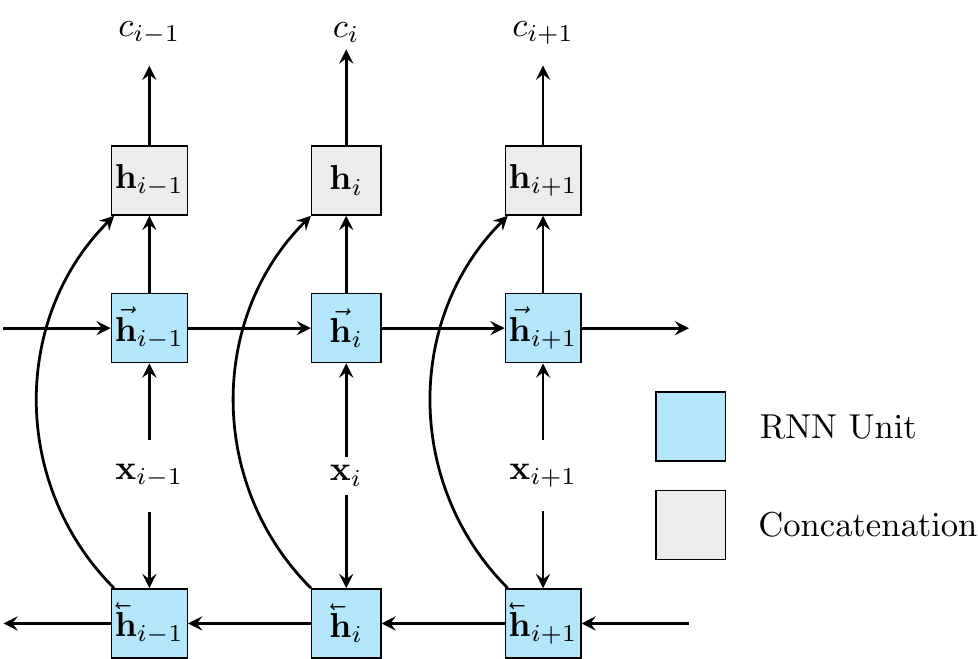}
    \caption{Bi-directional RNN for confidence prediction}
    \label{fig:bilstm}
\end{figure}

Recently there has been interest in examining modern forms of neural networks for confidence estimation. Figure~\ref{fig:bilstm} shows a bi-directional recurrent neural network (BiRNN) architecture examined in \cite{DelAgua2018aS,ragni2018confidence}.
Given a sequence of word level feature vectors ${\bm X}_{1:T}={\bm x}_{1},\ldots,{\bm x}_{T}$, the BiRNN makes use of {\em forward} and {\em backward} recurrent states to predict the sequence of confidence scores ${\bm c}_{1:T}=c_1,\ldots,c_T$. The recurrent states are aimed at encoding the complete past or future information respectively. In the simplest case the forward state is defined recursively by
\begin{equation}
\overrightarrow{\bm h}_{i} = {\bm\sigma}({\bm W}^{(\overrightarrow{h})} \overrightarrow{\bm h}_{i-1} + {\bm W}^{(x)} {\bm x}_{i})
\label{eq:birnn_hidden}
\end{equation}
where ${\bm W}^{(\overrightarrow{h})}$ and ${\bm W}^{(x)}$ are weight matrices for the forward state and the feature vector respectively, ${\bm\sigma}(\cdot)$ is an element-wise non-linearity, such as sigmoid, $\overrightarrow{\bm h}_{0}$ can be set to ${\bm 0}$ or learnt. The backward state can be defined analogously. Given the forward and backward states at time $i$, the confidence score can be predicted by
\begin{equation}
c_{i} = \sigma({\bm w}^{(c)^{\sf T}} {\bm h}_{i} + b^{(c)})
\label{eq:birnn_confidence}
\end{equation}
where ${\bm h}_{i}=[\overrightarrow{\bm h}_{i}^{\sf T}\;\overleftarrow{\bm h}_{i}^{\sf T}]^{\sf T}$, ${\bm w}^{(c)}$ and $b^{(c)}$ are weight vectors and a scalar bias, $\sigma$ is a non-linearity mapping confidence scores to a $[0,1]$ range. Note that unlike many other supervised learning problems, the targets for confidence scores need to be derived by automatically aligning predicted and manually transcribed word sequences \cite{Li_2019}. 

\begin{figure}[htbp]
    \centering
    \includegraphics[trim={0 20 0 0},width=7cm]{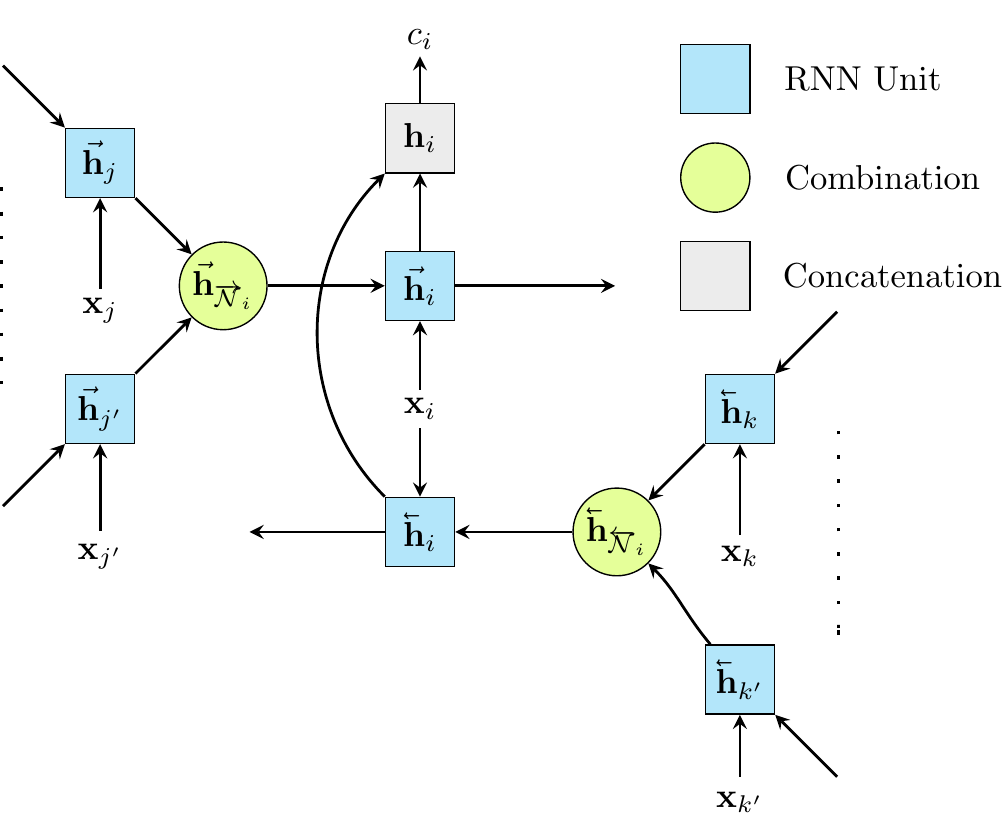}
    \caption{Bi-directional lattice RNN for confidence prediction}
    \label{fig:bilatlstm}
\end{figure}

The BiRNN is inherently limited to sequence data. As discussed in Section~\ref{sec:blackbox} one-best sequences carry only a small portion of information otherwise available in either constrained (CN) or unconstrained (lattice) DAG formats. Those DAGs are highly flexible structures that can be additionally enriched with a wide range of features \cite{Zweig2009aS, Ragni2011bS}. Recently there has been much interest in examining neural network extensions to DAGs and other general graph structures \cite{ladhak2016latticernn, Li_2019, Zhang2019aS}. The key question that any such approach needs to answer is how information associated with multiple graph arcs or nodes is combined. Figure~\ref{fig:bilatlstm} illustrates one such bi-directional approach for lattices (BiLatRNN). 
Compared to the sequence model in Figure~\ref{fig:bilstm}, the lattice model has one or more past recurrent states which propagates the current state to one or more subsequent states. In order to handle a variable number of past recurrent states, BiLatRNN makes use of an attention mechanism to create a combined representation
\begin{equation}
\overrightarrow{\bm h}_{\overrightarrow{\mathcal N}_{i}} = \sum_{j\in\overrightarrow{\mathcal N}_{i}} \alpha_{j} \overrightarrow{\bm h}_{j}
\label{eq:bilatrnn_aggregate}
\end{equation}
where $\overrightarrow{\mathcal{N}}_{i}$ is a set of incoming arcs for arc $i$. The attention mechanism makes use of arc contributions ${\bm e}$ to yield attention weights
\begin{equation}
\alpha_j = 
\exp(e_j)\bigg/\sum\limits_{j'\in{\overrightarrow{\mathcal N}_{i}}} \exp(e_{j'})
\label{eq:comb-weights}
\end{equation}
There are numerous ways for how arc contributions can be defined, such as scaled dot-product self-attention \cite{Vaswani2017aS}
\begin{equation}
e_{j} = \overrightarrow{\bm h}_{j}^{\sf T} \begin{bmatrix} \frac{1}{\sqrt{\dim(\overrightarrow{\bm h}_{j})}} {\bm I} \end{bmatrix} \overrightarrow{\bm h}_{j}
\end{equation}
multiplicative self-attention \cite{Luong_2015} with trainable weights ${\bm W}^{(m)}$,
\begin{equation}
e_{j} = \overrightarrow{\bm h}_{j}^{\sf T} {\bm W}^{(m)} \overrightarrow{\bm h}_{j}
\end{equation}
or additive attention \cite{Luong_2015, bahdanau2014neural} with trainable weights ${\bm W}^{(q)}$ and ${\bm w}^{(a)}$.
\begin{equation}
e_{j} = \sigma\left({\bm w}^{(a)^{\sf T}} {\bm\sigma}\left({\bm W}^{(q)} \begin{bmatrix}{\bm k}_{j}^{\sf T} & \overrightarrow{\bm h}_{j}^{\sf T}\end{bmatrix}^{\sf T}\right)\right)
\label{eq:latrnn_attention_additive}
\end{equation}
The additive form offers flexibility by ``querying" the state $\overrightarrow{\bm h}_{j}$ with a key ${\bm k}_{j}$. In previous work \cite{Li_2019} the key was set to
\begin{equation}
{\bm k}_{j} = \begin{bmatrix}
\log(\hat{c}_{j}) & \log(\hat{\mu}_{j}) & \log(\hat{\sigma}_{j})
\end{bmatrix}^{\sf T}
\end{equation}
where $\hat{c_j}$, $\hat{\mu}_{j}$ and $\hat{\sigma}_{j}$ are posterior probability, mean and standard deviation of all arc posterior probabilities which overlap in time with arc $j$. Such key design should enable the attention mechanism to downweight states of unlikely paths. Once the combined representation have been obtained, the current state can be updated by
\begin{equation}
\overrightarrow{\bm h}_{i} = {\bm\sigma}({\bm W}^{(\overrightarrow{h})} \overrightarrow{\bm h}_{\overrightarrow{\mathcal N}_{i}}+{\bm W}^{(x)} {\bm x}_{i})
\label{eq:bilatrnn_hidden_forward}
\end{equation}
The confidence score prediction is then done using equation~\eqref{eq:birnn_confidence}. The targets for lattice arc confidence scores are generated by extending the alignment algorithm for one-best sequences as described in \cite{Li_2019}. 

\section{Features}
\label{sec:featues}
As discussed in Section \ref{sec:blackbox}, a large amount of information is produced during the decoding process. However, for users of black box ASR typically only the one-best word sequence ${\bm w}_{1:T}=w_1,\ldots,w_{T}$ is available. If posterior probabilities and durations were also propagated the complete set of word level features could be expressed as
\begin{equation}
{\bm x}_{i}^{(w)} = \begin{bmatrix}
{\bm e}_{w_{i}}^{\sf T} & d_{w_i} & \log(\hat{c}_{w_i})
\end{bmatrix}^{\sf T}
\label{eq:word-features}
\end{equation}
where ${\bm e}_{w_{i}}$ is word $w_i$ represented as a one-hot encoding or embedding, $d_{w_i}$ is the word duration, and $\hat{c}_{w_i}$ is the posterior probability. The word embedding is a continuous word representation \cite{mikolov2013distributed} that can either be trained jointly with the rest of the neural network or independently on large quantities of text data \cite{Peters_2018, devlin2018bert} and then possibly fine-tuned. These simple features have been used with both BiRNN \cite{ragni2018confidence} and BiLatRNN \cite{Li_2019} for confidence prediction. 

\begin{figure}[htbp]
    \centering
    \includegraphics[width=6cm]{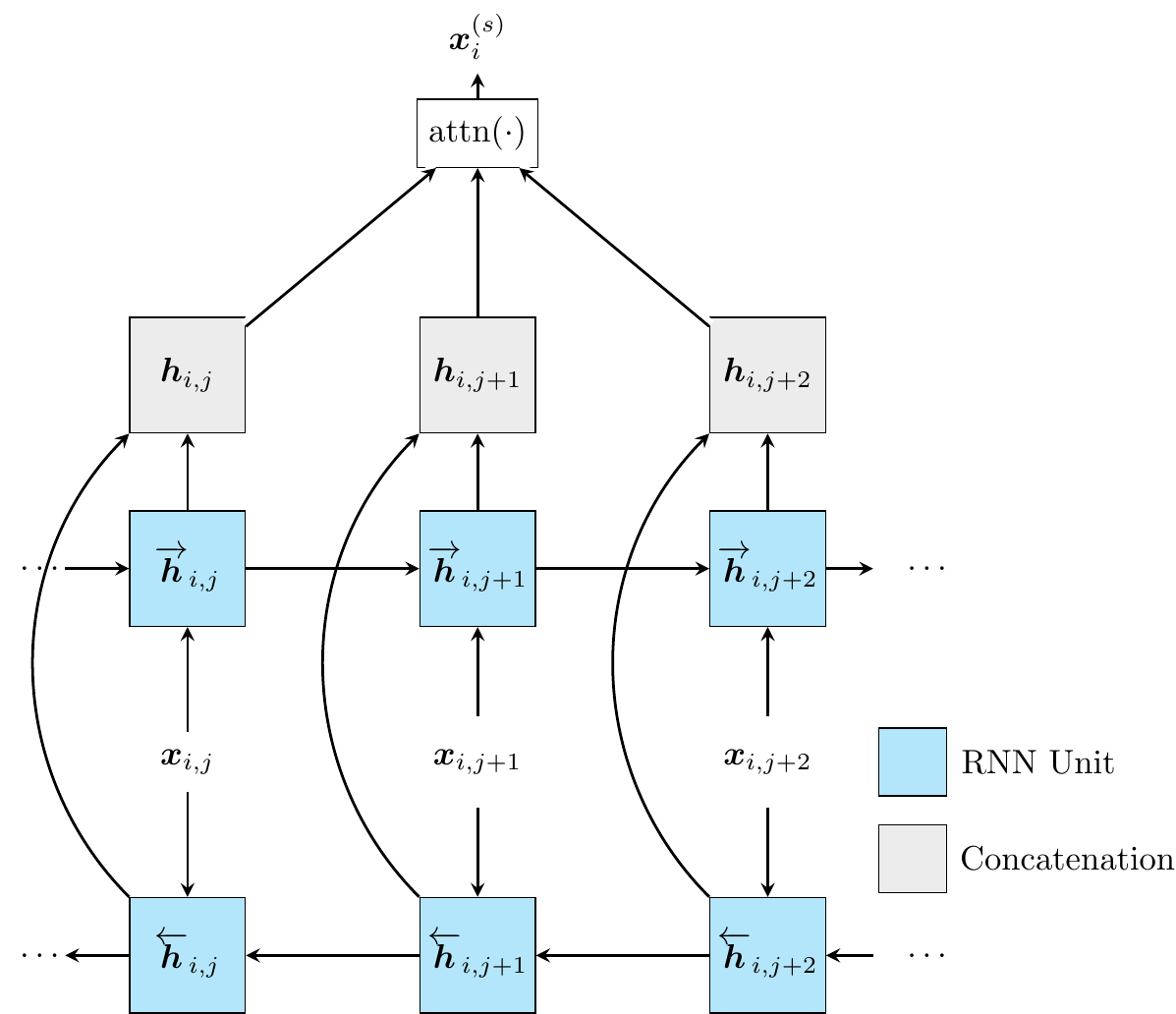}
    \caption{Bi-directional RNN sub-word encoder}
    \label{fig:birnnencoder}
\end{figure}

As mentioned in Section~\ref{sec:latrnn}, a wide range of additional information can be augmented to graph structures such as confusion networks and lattices. Any word level information can be added by simply extending the number of features in equation~\eqref{eq:word-features}. The use of sub-word information, such as phone, grapheme, morpheme, or byte-pair encoding, is more complicated due to variable length nature of sub-word sequences. A fixed length representation can be obtained by adopting the attention mechanism described in Section~\ref{sec:latrnn}
\begin{equation}
{\bm x}_{i}^{(s)} = \sum_{j\in{\mathcal S}_{i}} \alpha_{i,j} {\bm h}_{i,j}
\end{equation}
where ${\mathcal S}_{i}$ is a sequence of sub-word units for word $w_i$, $\alpha_{i,j}$ and ${\bm h}_{i,j}$ are an attention weight and continuous representation for sub-word unit $s_j$ respectively. There are several options for how sub-word representations ${\bm h}_{i,j}$ can be derived. In the simplest case, sub-word features can be defined in a similar manner to the word features by
\begin{equation}
{\bm x}_{i,j} = \begin{bmatrix}
{\bm\epsilon}_{s_j}^{\sf T} & d_{s_j} & \log(\hat{c}_{s_j})
\end{bmatrix}^{\sf T}
\end{equation}
where ${\bm\epsilon}_{s_j}$ is either a one-hot encoding or embedding, $d_{s_j}$ and $\hat{c}_{s_j}$ are duration and posterior probability respectively for sub-word $s_j$. A more powerful approach would be to use a bi-directional {\em encoder} as shown in Figure~\ref{fig:birnnencoder}.
To estimate sub-word attention weights ${\bm\alpha}_{i}$ for each word $w_i$ it is possible to use one of the approaches discussed in Section~\ref{sec:latrnn}. For instance, the use of additive attention in equation~\eqref{eq:latrnn_attention_additive} offers a number of interesting choices for selecting keys ${\bm k}_{i,j}$ to match against hidden states ${\bm h}_{i,j}$ that may include not only sub-word but also word level information. 

As discussed in Section \ref{sec:blackbox}, one-best word sequences are usually obtained from CNs rather than lattices, which makes information encoded into the latter not directly available for the former. Therefore, all prior work examined confidence estimation either based on CN or lattice output. Lattices, however, provide a rich and flexible framework for representing not only already available information but also various other external sources. As a result, they naturally facilitate interesting and often powerful features. Two of the simplest lattice features are the acoustic model score and the language model score \cite{Pinto2005ConfidenceMI}. More intricate features include acoustic stability \cite{Wessel2001aS} and hypothesis density \cite{Kemp97estimatingconfidence}. In order to make those features available, an alignment of CNs to lattices can be performed to match each CN arc to one or more lattice arcs depending on the time overlap tolerance specified. Provided that the lattices are large enough, the chance that any given CN arc cannot be matched to at least one lattice arc is small. Once lattice arcs have been aligned with CN arcs a number of approaches, such as simple averaging or an attention mechanism, can be used to yield lattice features for CN arcs.

\section{Experiments}
\label{sec:exp}
The experiments in this work were conducted on the decoded output from a graphemic ASR system trained for the IARPA OpenKWS 2016 competition \cite{Ragni2017stimulatedtraining}. The audio recordings consist of Georgian conversational telephone speech, of which 25 hours was used for training and testing confidence estimation approaches. The predictions from this system, which are treated as a black box, are split into independent training, cross-validation, and test sets with an $8:1:1$ ratio. After CN decoding, the ASR system has a word error rate of approximately 38\%, which leads to an imbalanced distribution of correct and incorrect word predictions. All models, BiRNN and BiLatRNN, use a single 128-dimensional bi-directional LSTM layer with a fully connected hidden layer consisting of 128 neurons. The sub-word encoder uses a similar architecture based on a 10-dimensional GRU layer. The results are presented in terms of two standard metrics, normalised cross-entropy (NCE) \cite{ragni2018confidence,siu1997improved} and area under the curve (AUC), where the precision-recall curve is used to mitigate the effect of the dataset imbalance \cite{davis2006relationship}. The AUC improvements have a slight tendency to favour the low recall high precision area of the curve. The targets were obtained through a Levenshtein type alignment of predicted words to the reference transcription. A random classifier in this setup will render an AUC score of 0.6310.

The set of word level features used include a 50-dimensional {\footnotesize\tt fastText} \cite{bojanowski2017enriching} pre-trained word embedding, duration, and posterior without and with decision tree mapping \cite{Evermann2000aL}. Table \ref{tab:word-features} demonstrates how the incremental addition of these input features to the BiRNN model results in performance improvements relative to the use of simple word embeddings.
\begin{table}[htbp]
\centering
\begin{tabular}{l||cc}
\hline
Word Features & NCE & AUC\\
\hline\hline
words & 0.0358 & 0.7496 \\
\;\;\;+duration &  0.0541 & 0.7670 \\
\;\;\;\;\;\;+ posteriors & 0.2765 & 0.9033\\
\;\;\;\;\;\;\;\;\; + mapping & \textbf{0.2911} & \textbf{0.9121}\\
\hline
\end{tabular}
\caption{Confidence estimation performance using word features}
\label{tab:word-features}
\end{table}
As expected, the use of words and durations yields low, although higher than random, AUC values whilst the introduction of posteriors sees a large performance improvement. 

The set of sub-word level features used included a 4-dimensional {\footnotesize\tt word2vec} \cite{mikolov2013distributed}  pre-trained grapheme embedding and duration. As described in Section \ref{sec:featues}, sub-word features can be incorporated into word level models using an attention mechanism applied either directly to sub-word features or to encoder states. A comparison of attention approaches (see Section~\ref{sec:latrnn}), which is not reported here due to space constraints, showed that the additive attention with the sub-word embedding and duration as a key yielded slightly better results and hence is used in the rest of this section. 
\begin{table}[htbp]
\centering
\begin{tabular}{l||cc}
\hline
Sub-word Features & NCE & AUC\\
\hline\hline
none & 0.2911 & 0.9121\\
embedding & 0.2936 & 0.9127 \\
\;\;\;+ duration & 0.2944 & 0.9129\\
\;\;\;\;\;\;+encoder & \textbf{0.2978} & \textbf{0.9139}\\
\hline
\end{tabular}
\caption{Impact of sub-word features}
\label{tab:subword-features}
\end{table}

Table~\ref{tab:subword-features} shows that the use of sub-word information (embedding, +duration) and more complex representations (+encoder) yields small but consistent gains. 

The BiRNN examined so far lacked any information about competing transcriptions available within CNs. Depending on application there are several ways how such information can be utilised. If the task is to predict confidence scores for one-best word sequences (as in this work) the training loss should be accumulated over one-best sequences only. However, if confidence scores of all arcs are of interest (as in previous work \cite{Li_2019}) the training loss should be accumulated over all arcs. Note that the forward propagation is done through all arcs irrespective of the choice made above.
\begin{table}[htbp]
\centering
\begin{tabular}{cc||cc}
\hline
Confusions & Loss & NCE & AUC\\
\hline\hline
$1$-best & $1$-best & 0.2911 & 0.9121\\
CN & $1$-best & 0.2931 & 0.9201\\
CN & CN & 0.2934 & 0.9178\\
\hline
\end{tabular}
\caption{Impact of word confusion information}
\label{tab:confusion}
\end{table}
Table~\ref{tab:confusion} shows that although both BiLatRNN approaches yield gains over the one-best baseline, the former as expected yields better AUC values. Table~\ref{tab:lattice} also shows that word confusion and sub-word information are quite complimentary, yielding significant gains over word only one-best baseline.  
\begin{table}[htbp]
\centering
\begin{tabular}{l||cc}
\hline
\hspace{0.4cm}Features & NCE & AUC\\
\hline\hline
word (all) & 0.2911 & 0.9121\\
\;\;\;+confusions & 0.2934 & 0.9201\\
\;\;\;\;\;\;+sub-word & 0.2998 & 0.9228\\
\;\;\;\;\;\;\;\;\;+lattice & \textbf{0.3004} & \textbf{0.9231}\\
\hline
\end{tabular}
\caption{Impact of word confusion, sub-word and lattice features}
\label{tab:lattice}
\end{table}
As discussed in Section~\ref{sec:featues}, a range of lattice features can be incorporated into BiLatRNNs by aligning lattices to CNs. As a proof of concept this work examined a simple set of lattice features: the acoustic and language model scores. Given a relatively large set of lattices and a tight threshold on time overlap, the alignment process failed to match 1.7\% of training utterances, which, in this work, were removed from training. Note that significantly larger lattices can be obtained by simply increasing decoding beam size. Table~\ref{tab:lattice} shows that the BiLatRNN can leverage even the simplest of lattice features with more gains expected from more complex approaches.



\begin{samepage}
\section{Conclusion}
\label{sec:conclusion}
With black box automatic speech recognition (ASR) systems becoming more popular, the importance of error mitigation strategies grows. Despite clear evidence from the literature that word sequences alone are not adequate for building accurate applications, the restricted form of one-best remains the {\em de facto} standard output of commercial ASR. Word sequences, however, provide a limited opportunity for devising even the simplest error mitigation strategy, a confidence score. This paper examines a hypothetical scenario where progressively more (normally discarded) information, such as confusion networks and lattices, are propagated to the user. To leverage these graph structures a bi-directional lattice recurrent neural network was used and extended to handle sub-word information. Experimental results on the challenging IARPA OpenKWS 2016 task show that additional information is crucial and can be easily leveraged using available neural network approaches
\end{samepage}

\newpage
\bibliographystyle{IEEEbib}
\bibliography{strings,refs}

\end{document}